\documentclass[preprint,showpacs,preprintnumbers,amsmath,amssymb]{revtex4}

\usepackage{graphicx}
\usepackage{dcolumn}
\usepackage{bm}
\usepackage{mathrsfs}

\begin{document}

\title{Color-flavor locked strange matter and strangelets at finite temperature}

\author{L. Paulucci}
 \email{paulucci@fma.if.usp.br}
\affiliation{Instituto de F\'\i sica - Universidade de S\~ao Paulo\\
Rua do Mat\~ao, Travessa R, 187, 05508-090, Cidade Universit\'aria\\
S\~ao Paulo SP, Brazil}

\author{J. E. Horvath}
\affiliation{Instituto de Astronomia, Geof\'\i sica e Ci\^encias
Atmosf\'ericas - Universidade de S\~ao Paulo\\
Rua do Mat\~ao, 1226, 05508-900, Cidade Universit\'aria\\
S\~ao Paulo SP, Brazil}

\date{\today}

\begin{abstract}
It is possible that a system composed of up, down and strange
quarks consists the true ground state of nuclear matter at high
densities and low temperatures. This exotic plasma, called strange
quark matter (SQM), seems to be even more favorable energetically
if quarks are in a superconducting state, the so-called
color-flavor locked state. Here are presented calculations made on
the basis of the MIT bag model considering the influence of finite
temperature on the allowed parameters characterizing the system
for stability of bulk SQM (the so-called stability windows) and
also for strangelets, small lumps of SQM, both in the color-flavor
locking scenario. We compare these results with the unpaired SQM
and also briefly discuss some astrophysical implications of them.
Also, the issue of strangelet's electric charge is discussed. The
effects of dynamical screening, though important for non-paired
SQM strangelets, are not relevant when considering pairing among
all three flavor and colors of quarks.
\end{abstract}



\maketitle

\section{Introduction}

Three decades ago it was proposed that systems composed of an
unconfined Fermi liquid of up, down, and strange quarks could be
absolutely stable \cite{Wit, Bodmer, Chin, Terazawa}. In its
simplest Fermi liquid picture, stability depends on whether it
would be possible or not to lower the energy of a system composed
of quarks $u$ and $d$ by converting (through weak interactions)
approximately one third of its components into the much more
massive strange quark due to the introduction of a third Fermi
sea.

Within the well-known MIT Bag Model \cite{MIT}, it has been shown
that this stability may be realized for a wide range of parameters
of strange quark matter (SQM) in bulk \cite{Farhi84}.
Other calculations also indicate that SQM can be absolutely stable
within different frameworks, e.g. shell model \cite{Jaf}.

The attractive force between quarks that are antisymmetric in
color tend to make quarks near the Fermi surface to pair at high
densities, breaking the color gauge symmetry, and causing the
phenomena of color superconductivity. Recently, studies have
indicated that the color-flavor locked (CFL) state, in which all
the quark modes are gaped, seems to be even more favorable
energetically, widening the stability window \cite{CFL1, CFL2,
CFL3, Madsen01, Lugones}.

If strange quark matter or CFL matter is indeed the ground state
of cold and dense baryonic matter there would be some important
astrophysical implications. For instance, all neutron stars would
actually have their interiors composed only of exotic matter
\cite{SS1, SS2, SS3, Mac, SS5, SS6}, see also \cite{Madsen04,
Weber05, Xu07} for recent reviews. The existence of strange stars
would likely imply the presence of strangelets among cosmic ray
primaries. A few injection scenarios have been considered as
likely sites: the merging of compact stars (though not addressed
in full detail yet) \cite{Mergers, Rosinska}, strange 
matter formation in type II supernova
\cite{Mac}, and acceleration from strange pulsars
\cite{ChengUsov}. Several cosmic ray events have been tentatively
identified with primary strangelets (mainly the Centauro and Price
events, and more recently data from the HECRO-81, ET event, and
AMS01 experiments \cite{ET, Price, Data1, Data2, Data3, Boiko})
for the data obtained indicate a high penetration of the particle
in the atmosphere, low charge-to-mass ratio, and exotic
secondaries. New experiments are being designed that could
identify these exotic primaries with the purpose to definitively
test the validity of the Bodmer-Witten-Terazawa conjecture
\cite{AMS, AMS2, Sandweiss}.

For the description of these \textit{finite} size lumps of strange
matter, (termed \textit{strangelets}) a few terms have to be added
to the bulk one in the free energy (see \cite{Madsen98} and
\cite{Mads} for details). Large lumps will have essentially the
same structure than bulk matter, with a small depletion of the
massive strange quarks near the surface resulting in a net
positive charge, a feature also expected for smaller chunks
\cite{Jaf},\cite{Mads} which thus resemble heavy nuclei.

Strangelets without pairing at finite temperature have been first
analyzed by Madsen \cite{Madsen98} in the $m_s=0$ approximation. A
more complete description has been given by He et al.
\cite{Chineses}, in which energy, radius, electric charge
(unscreened), strangeness fraction and minimum baryon number were
presented.

CFL strangelets at $T=0$ were discussed in Refs. \cite{Madsen01,
Peng}. More recently, a finite-temperature analysis using pertubative
QCD appeared \cite{Schmitt}. We address in this paper
the issues of surface and curvature energies at $T > 0$, which are
potentially important for fragmentation of CFL SQM in
astrophysical environments among other things.

This paper in structured as follows: in section II we describe the
theoretical approach used to determine the parameters
characterizing CFL SQM at finite temperature for the construction
of the windows of stability in bulk and strangelets; in section
III, we present the numerical results for CFL SQM and compare then
with unpaired SQM; in section IV, we present our final discussion
and conclusions.

 \section{Windows of stability}

\subsection{Bulk matter}

Unpaired SQM in bulk contains $u$, $d$, $s$ quarks and also
electrons to maintain charge neutrality. The chemical balance is
maintained by weak interactions and neutrinos assumed to escape
from the system. If SQM is in a CFL state in which quarks of all
flavors and colors near the Fermi surface form pairs, an equal
number of flavors is enforced by symmetry and the mixture is
automatically neutral \cite{Rajagopal2001}. In this case, the
condition is that the \textit{Fermi momentum} for the three quarks
are equal, so that $3\mu=\mu_u+\mu_d+\mu_s$ and the common Fermi
momentum is $\nu=2\mu-(\mu^2+m_s^2 / 3)^{1/2}$.

For bulk CFL SQM, the thermodynamical potential for the system to order
$\Delta^2$ is \cite{Rajagopal2001, AlfordReddy}

\begin{equation}
\Omega_{CFL}=\sum_i\Omega_i-\frac{3}{\pi^2}\Delta^2\mu^2+B
\end{equation}
\\
where $\Delta$ is the pairing energy gap and the term associated with
this parameter is the binding energy of the diquark condensate. The term
$\Omega_{free}=\sum_i\Omega_i$, mimics an unpaired state where all quarks
have a common Fermi momentum, and $i$ stands for quarks $u,d,s$ and
gluons (there is no electrons in the CFL state).

On the basis of the MIT bag model with $\alpha_c$=0 \footnote{When
considering a finite strong coupling constant, $\alpha_c$, it has
been shown that it can in fact correspond to an effective
reduction on the value of the MIT bag constant $B$
\cite{Farhi84}.}, the thermodynamic bulk potentials for each
component of the unpaired ``toy model'' system are given by

\begin{equation}\label{Omega}
\Omega_i=\mp T \int^{\infty}_{0}dk g_i \frac{k^2}{2\pi^2}\ln\Big[1\pm\exp\Big(-\frac{\epsilon_i(k)-\mu_i}{T}\Big)\Big]
\end{equation}
\\
where the upper sign corresponds to fermions and the lower to
bosons, $\mu$ and $T$ are the chemical potential and temperature,
$k$ and $\epsilon_i$ the momentum and energy of the particle,
respectively, and the factor $g_i$ is the statistical weight for
quarks and gluons (6 for quarks and antiquarks and 16 for gluons).
The limit of expression \ref{Omega} for $T\rightarrow 0$ is the
one given in \cite{Farhi84}, when the integral is made for momenta
ranging from zero to the Fermi one, since the Fermi-Dirac
distribution at $T=0$ for the unpaired state presents a sharp
cutoff at the Fermi momentum \footnote{The CFL state does not
actually present a sharp Fermi surface. See \cite{Alford} for more
details} being useless to perform the integration to $k\rightarrow
\infty$. At finite temperature, however, the broadening of the
Fermi-Dirac distribution occurs, hence the integration has to be
extended as well.

With these quantities we obtain the particle density given by
$n_i=-\partial \Omega_i /\partial \mu_i$ (which accounts for the influence
of the pairing condensate binding energy), and the total energy density,
$E=\sum_{i}(\Omega_i+n_i\mu_i)-3\Delta^2\mu^2/\pi^2+B+TS$, where
$S=-(\partial\Omega/\partial T)_{V,\mu}$ is the entropy.

In spite that most of the analysis is performed using a constant
value for $\Delta = \Delta_{0}$, the pairing gap is actually
dependent of the temperature of the system. Following the studies
of superconductivity in quark matter \cite{Schmitt, CFL3}, we used
for this dependence 

\begin{eqnarray}
\Delta(T)=2^{-1/3}\Delta_0\sqrt{1-\Big(\frac{T}{T_c}\Big)^2}\\
T_c=0.57\Delta(T=0)\times 2^{1/3}\equiv 2^{1/3}\times 0.57\Delta_0
\end{eqnarray}
\\
where $T_c$ is the critical temperature of the superconducting
system, above which the system can no longer support pairing
between quarks.

 \subsection{Strangelets}

As stated above, for the description of strangelets, it is
necessary to add surface and curvature contributions to the
thermodynamical potentials of bulk matter

\begin{equation}
\Omega_i=\mp T \int^{\infty}_{0}dk \frac{dN_i}{dk} \ln\Big[1\pm\exp\Big(-\frac{\epsilon_i(k)-\mu_i}{T}\Big)\Big]
\end{equation}

In the multiple reflection expansion \cite{Balian2, Hansson}
the density of states is given by

\begin{equation}
\frac{dN_i}{dk}=g_i\Big\{\frac{1}{2\pi^2}k^2\mathscr{V}+f^{(i)}_S\Big(\frac{m_i}{k}\Big)k\mathscr{S}+f^{(i)}_C\Big(\frac{m_i}{k}\Big)\mathscr{C}\Big\}
\end{equation}
\\
where $\mathscr{V}$, $\mathscr{S}$ and $\mathscr{C}$ stand for the 
volume, surface area and curvature of the strangelet, respectively.

The surface term for quarks is given by \cite{Berger}

\begin{equation}
f^{(q)}_S\Big(\frac{m_q}{k}\Big)=-\frac{1}{8\pi}\Big[1-\frac{2}{\pi}\arctan\Big(\frac{k}{m_q}\Big)\Big]
\end{equation}

For the curvature contribution, the following \textit{ansatz}
\cite{Mad94} for massive quarks is adopted

\begin{equation}
f^{(q)}_C\Big(\frac{m_q}{k}\Big)=\frac{1}{12\pi^2}\Big\{1-\frac{3k}{2 m_q}\Big[\frac{2}{\pi}-\arctan\Big(\frac{k}{m_q}\Big)\Big]\Big\}
\end{equation}
\\
while for gluons \cite{Balian}, $f^{(g)}_C=-1/ 6\pi^2$.

The energy is obtained as
$E=\sum_i(\Omega_i+N_i\mu_i)-3\Delta^2\mu^2/\pi^2V+BV+TS$ and
the mechanical equilibrium condition for a strangelet with vacuum
outside is given by $B=-\sum_i\partial\Omega_i / \partial V$. The
relation obtained for strangelets without pairing \cite{Chineses},
$\mu_u=\mu_d=\mu_s$, found by minimizing the free energy with
respect to the net number of quarks of each species and subjected
to the constraint $A=1/3 \sum_i N_i$, is here substituted by
$\mu_u=\mu_d$ and $\mu_s=\sqrt{\mu_u^2+m_s^2}$ which is actually a
second constraint imposed for pairing to hold. The value of the
common chemical potential is then obtained numerically by imposing
the mechanical equilibrium condition at a given set of $B$,
$\Delta$ and $m_s$.

The issue of Debye screening of the electric charge for
strangelets without pairing is of major importance in determining
the total charge of these particles \cite{Heiselberg}. In the
expression for energy density, there is a term proportional to
$A_0^2/\lambda_D^2$, where $A_0$ is the gauge field for the
massless gauge boson and $\lambda_D$ is the Debye screening
length. In this way, the general expression of the Debye screening
length may be written as

\begin{equation}\label{screen}
\lambda_D^{-2}\propto \frac{\partial^2 \text{Energy density}}{\partial\mu_e^2}
\end{equation}
\\
where $\mu_e$ is the chemical potential for electric charge. It means that
$\lambda_D$ is related to the response of a medium to a change in $\mu_e$.

In CFL matter the massless gauge boson is the rotated or $\Tilde{Q}$
photon and the \~Q-charge of all the Cooper pairs
forming the condensate is zero. Since all quasiparticles are gapped,
which is due the unbroken \~U(1)$_{em}$ gauge symmetry in the ground state,
the CFL phase is not an electromagnetic superconductor but a
\~Q-insulator \cite{CFL1}.

In CFL matter, the relevant electric chemical potential is
$\mu_{\Tilde{Q}}$ and the root mean square of (\ref{screen}) is zero
\cite{Alfordpers}, therefore the electric field is not screened
and the charge of a strangelet in the CFL state will be defined by
finite size effects only.

\section{Numerical results}

The so-called ``windows of stability'' (regions in the plane
$m_s-B$) for CFL matter in the framework of the MIT Bag Model are
shown in Fig. \ref{janela}. The minimum value for $B^{1/4}$ is
$145$ MeV, because a lower $B$ would cause the spontaneous decay
to non-strange matter ($u$ and $d$ quarks). As expected, the
matter becomes less bound at finite temperature as can be seen
both in the constant pairing gap approximation and when this
parameter is temperature dependent.

The influence of considering the more realistic case of $\Delta$
depending on the temperature, $\Delta=\Delta(T)$, shows an
destabilization of the system due to an effective reduction 
in the gap parameter. 
This conclusion holds even if the system is quite close to the
critical temperature for pairing between quarks. The system, then,
tends to approach the curves for $\Delta=0$. But it does not
exactly match the curves for $\Delta=0$ due  to the existence
of the extra term in the entropy 
($\partial [-3\mu^2/\pi^2\Delta(T)^2]/\partial T \simeq 3\mu^2/\pi^2T_c/0.41$,
when $T=T_c$) for the temperature dependent scenario.

\begin{widetext}

\begin{center}
\begin{figure}
\includegraphics[width=0.45\textwidth]{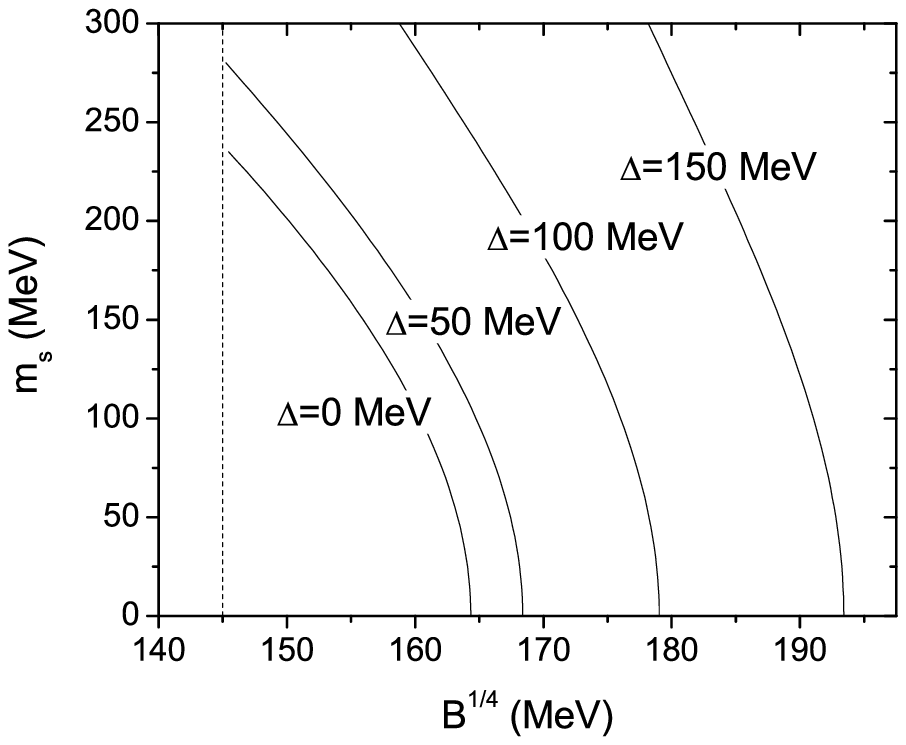}
\includegraphics[width=0.45\textwidth]{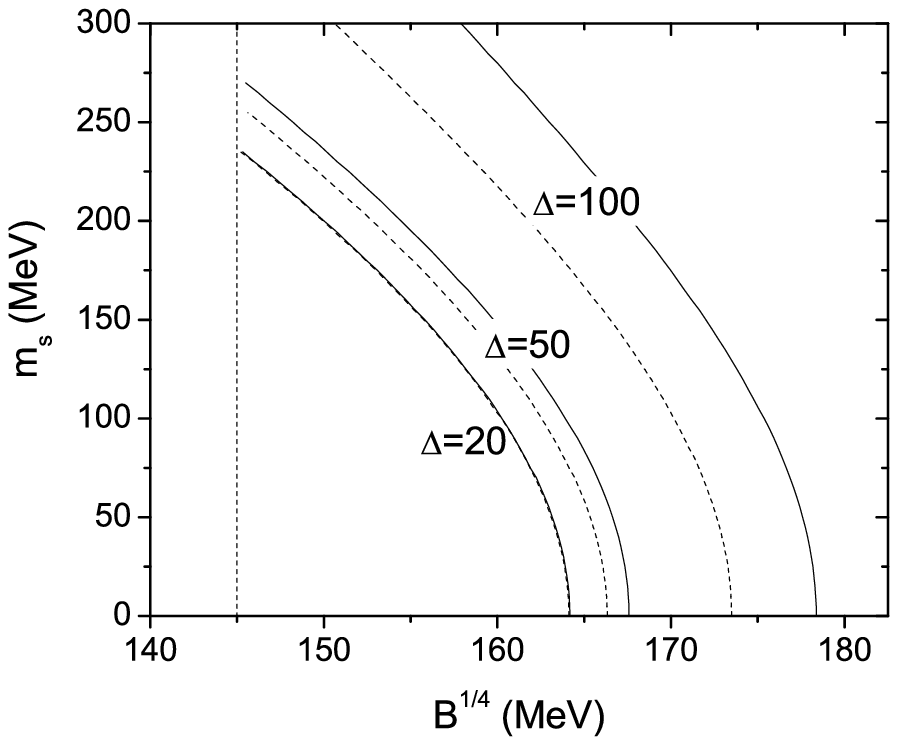}
\includegraphics[width=0.45\textwidth]{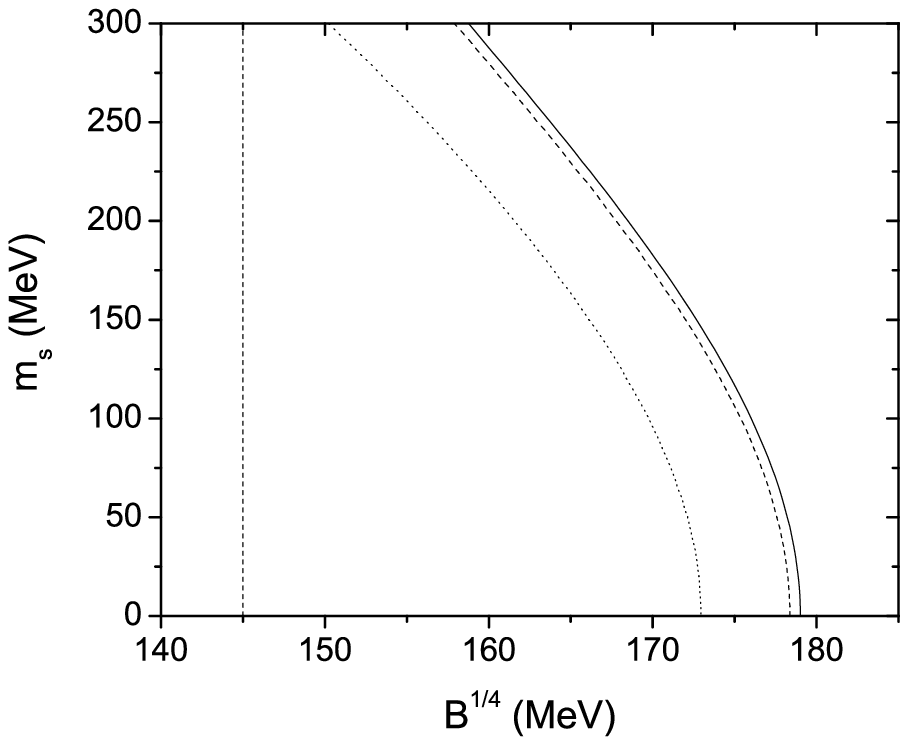}
\includegraphics[width=0.45\textwidth]{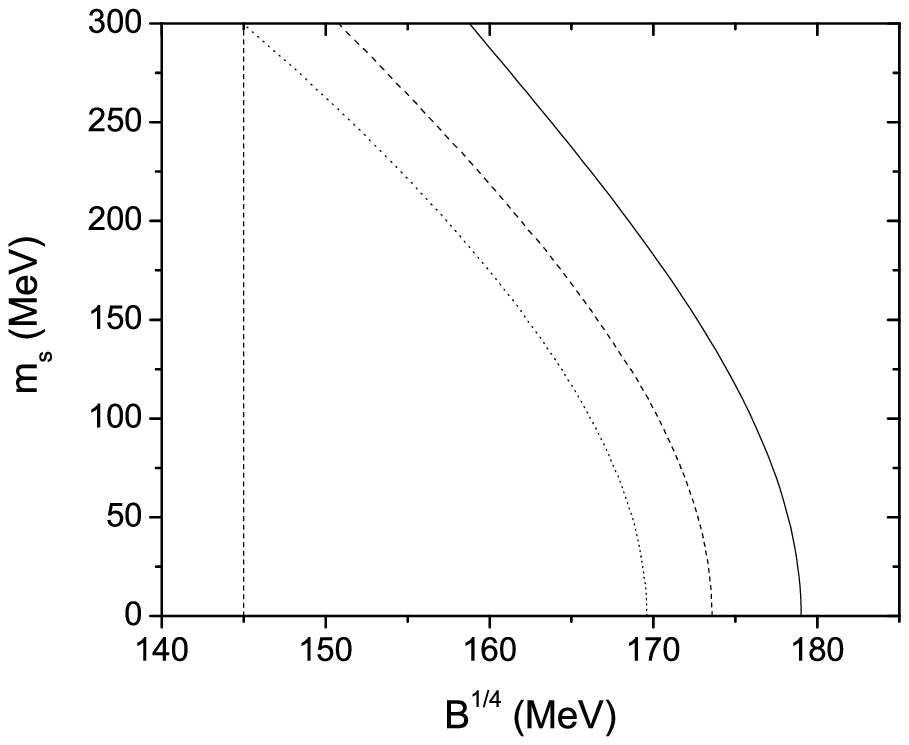}
\caption{Stability windows, regions bounded
by the vertical line at $B^{1/4}=145$ MeV and the curves of
$E/A=939$ MeV (shown
for different temperatures and $\Delta$), for CFL SQM. On the top
left panel, at $T=0$ (following reference \cite{Lugones}) and on 
the top right panel, for CFL SQM at T=10 MeV and values of
$\Delta$ as indicated. The full lines represent the calculations
made considering $\Delta$ constant with temperature and the dashed
lines for $\Delta=\Delta(T)$ (refer to text for details). On the
bottom panels, stability window for CFL SQM at finite temperature
and $\Delta=100$ MeV. The solid line is for null temperature, the
dashed line for $T=10$ MeV, and the dotted line, $T=30$ MeV. On
the left, the curves where obtained considering a fixed $\Delta$
and on the right, $\Delta=\Delta(T)$. All the curves presented are
calculated for fixed $E/A=939$ MeV labelled with the corresponding
value of $\Delta$, when necessary. The vertical line
is the minimum $B$ value for stability.}\label{janela}
\end{figure}
\end{center}

\end{widetext}

\begin{center}
\begin{figure}
\includegraphics[width=0.45\textwidth]{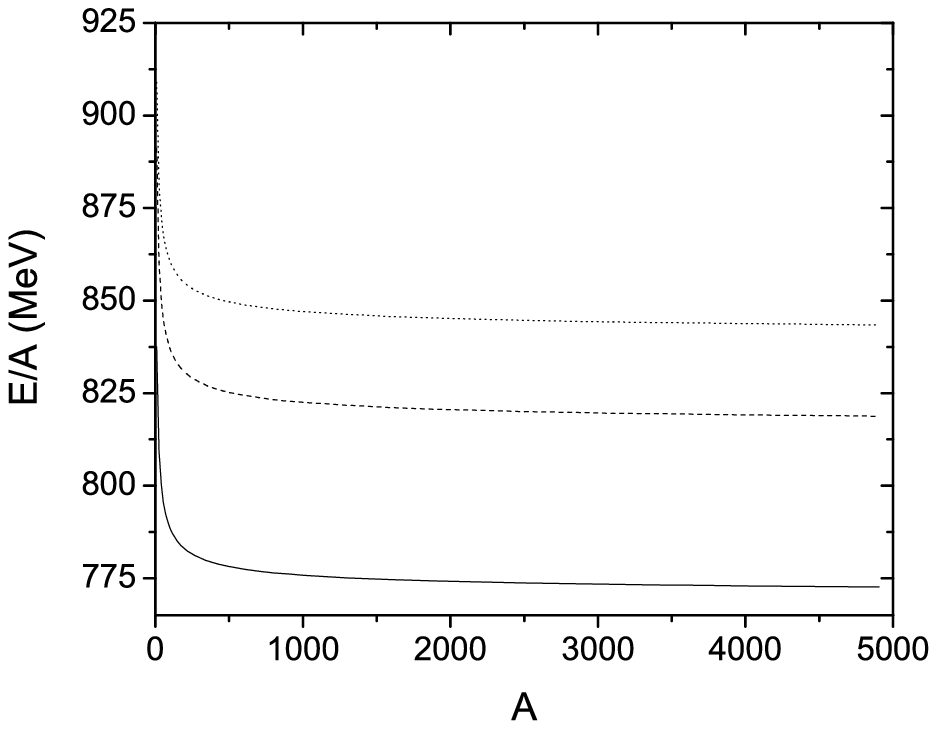}
\caption{Energy per baryon number of CFL strangelets as 
a function of $A$ calculated for
$\Delta=\Delta(T)$. The value of the fixed constants are $T=0$
(full curve), $T=15$ MeV (dashed curves), and $T=30$ MeV
(dotted curve), $m_s=150$ MeV,
$B^{1/4}=145$ MeV, and $\Delta=100$ MeV.}\label{energiaT}
\end{figure}
\end{center}

We have also calculated the structure of spherical strangelets,
numerically, with the results shown in Figs. \ref{energia} and
\ref{energiaT} for the total energy of these particles as a
function of different parameters characterizing them.

Just as in the case of bulk matter, there is a competition when
considering $\Delta=\Delta(T)$ between the lowering of the
effective pairing parameter and the raising of the chemical
potentials in the CFL quark matter and the extra term in the volumetric
entropy when compared to the case of a constant pairing gap parameter.
For finite size drops of SQM, the additional terms
of surface and curvature contributing in the thermodynamic potential
with opposite sign to the volumetric term are affected only by the
changes in $\mu$ (higher in the $\Delta(T)$ case) and on the strangelet
radius (lower but not significantly affected, the difference being less
than 1\%). The overall result is that the stability for a given set
of $m_s$, $B$, $\Delta_0$, $A$, and temperature is disfavored
in the dependent delta scenario, as it is in the bulk case.

\begin{widetext}

\begin{center}
\begin{figure}
\includegraphics[width=0.45\textwidth]{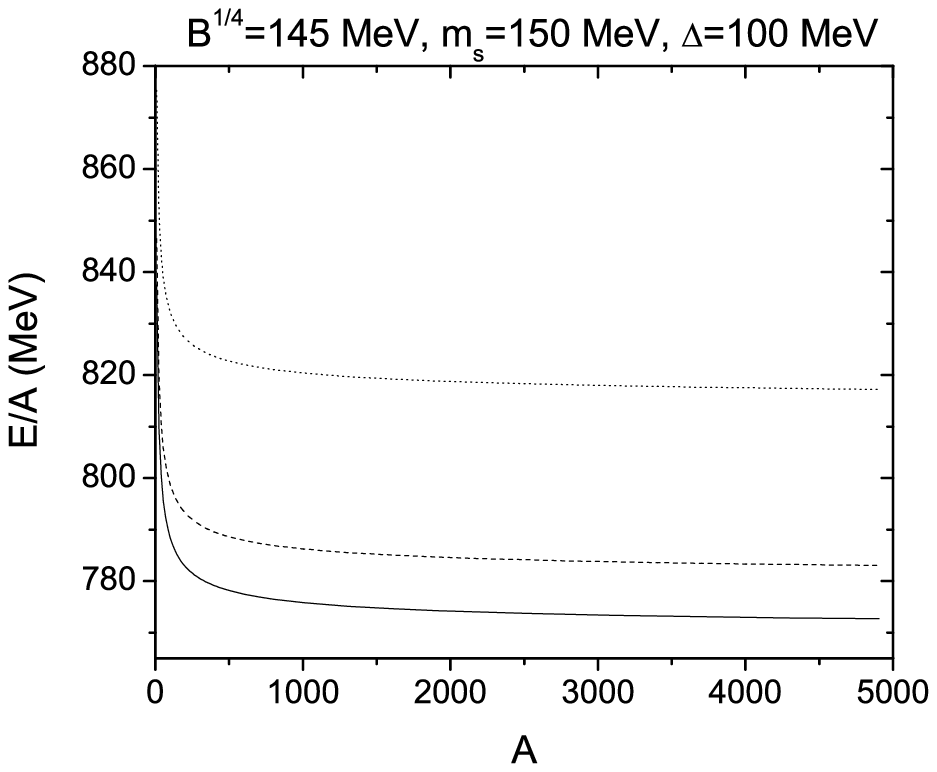}
\includegraphics[width=0.45\textwidth]{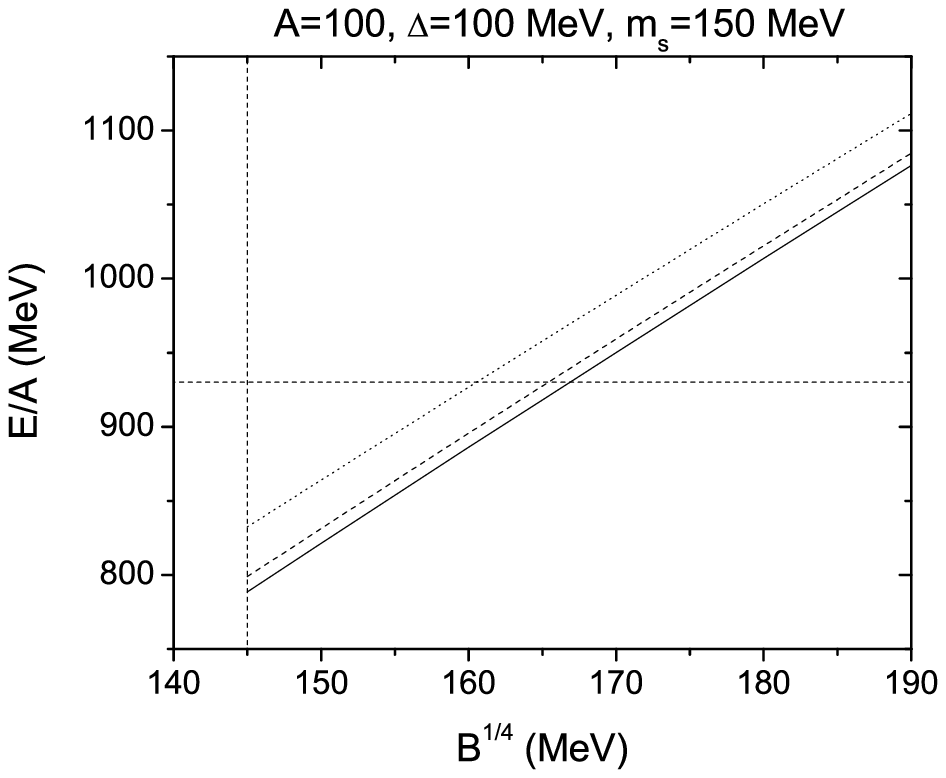}
\includegraphics[width=0.45\textwidth]{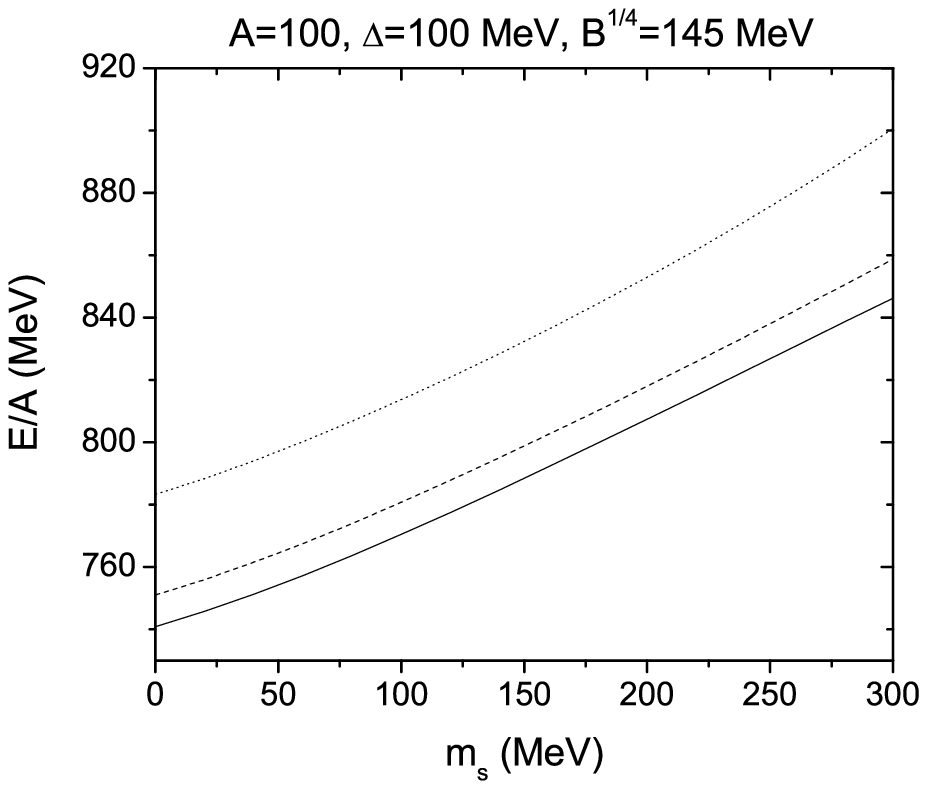}
\includegraphics[width=0.45\textwidth]{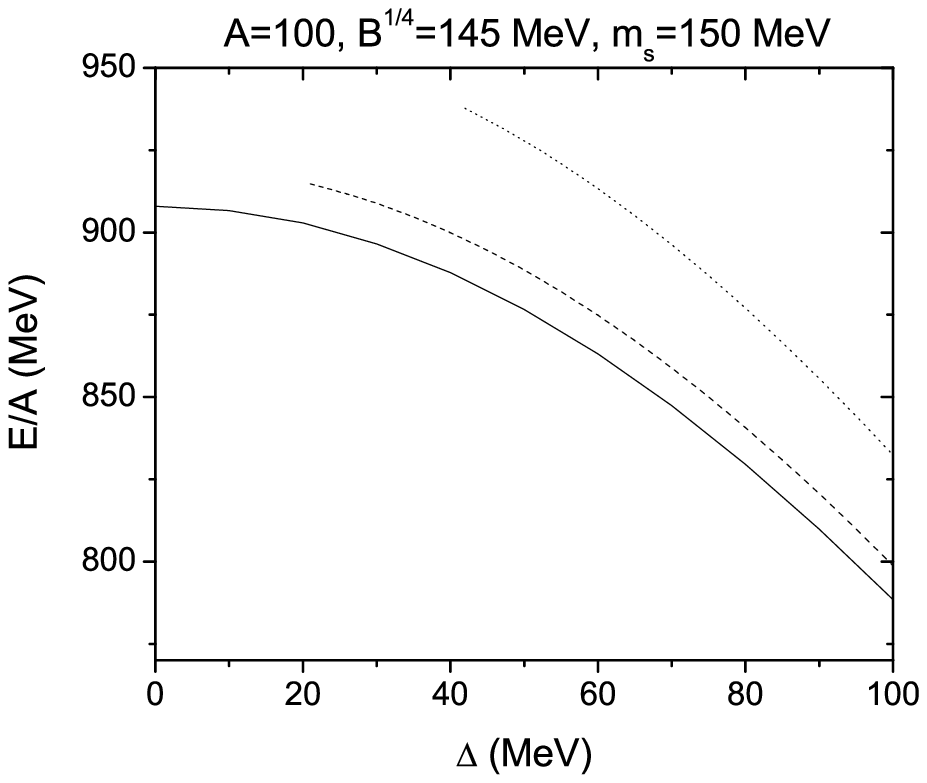}
\includegraphics[width=0.45\textwidth]{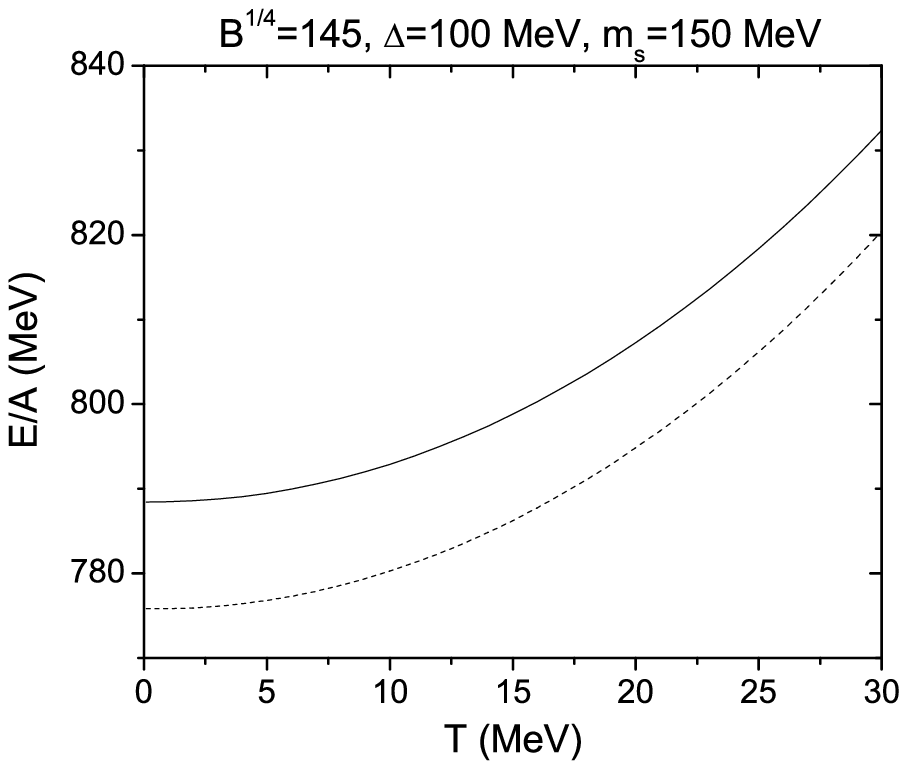}
\caption{Energy per baryon number as a function of the baryonic
number, bag constant, strange quark mass, pairing energy gap, and
temperature, from left to right, top to bottom, respectively. The
values of the fixed constants are indicated for each plot. The
first four plots are performed for $T=0$ (full curves), $T=15$ MeV
(dashed curves) and $T=30$ MeV (dotted curves). The last plot is
performed for $A=100$ (full curve) and $A=1000$ (dashed
curve).}\label{energia}
\end{figure}
\end{center}

\begin{center}
\begin{figure}
\includegraphics[width=0.45\textwidth]{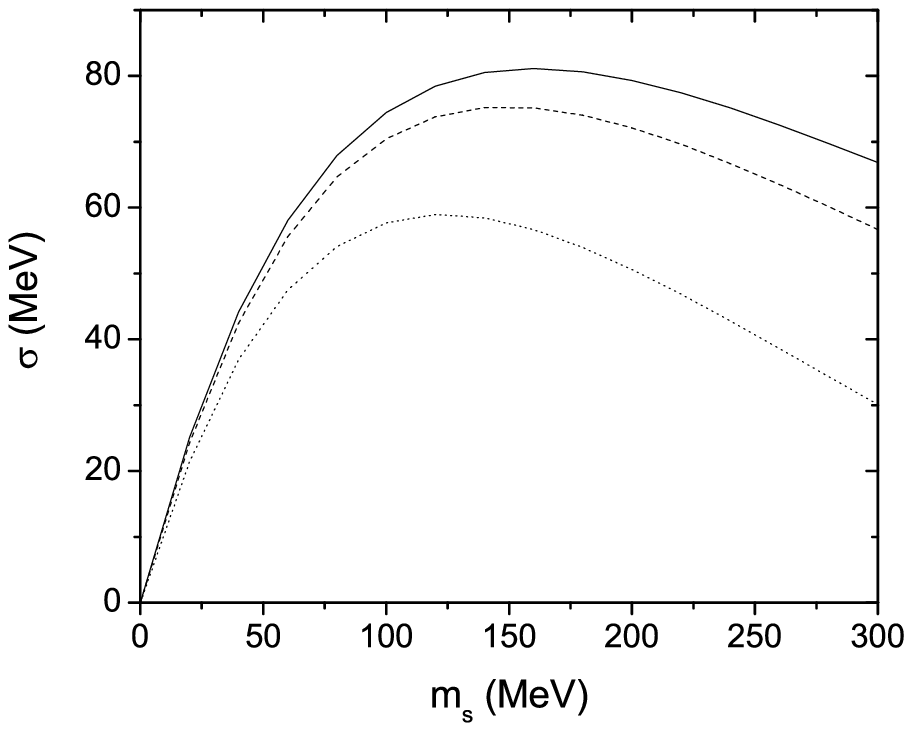}
\includegraphics[width=0.45\textwidth]{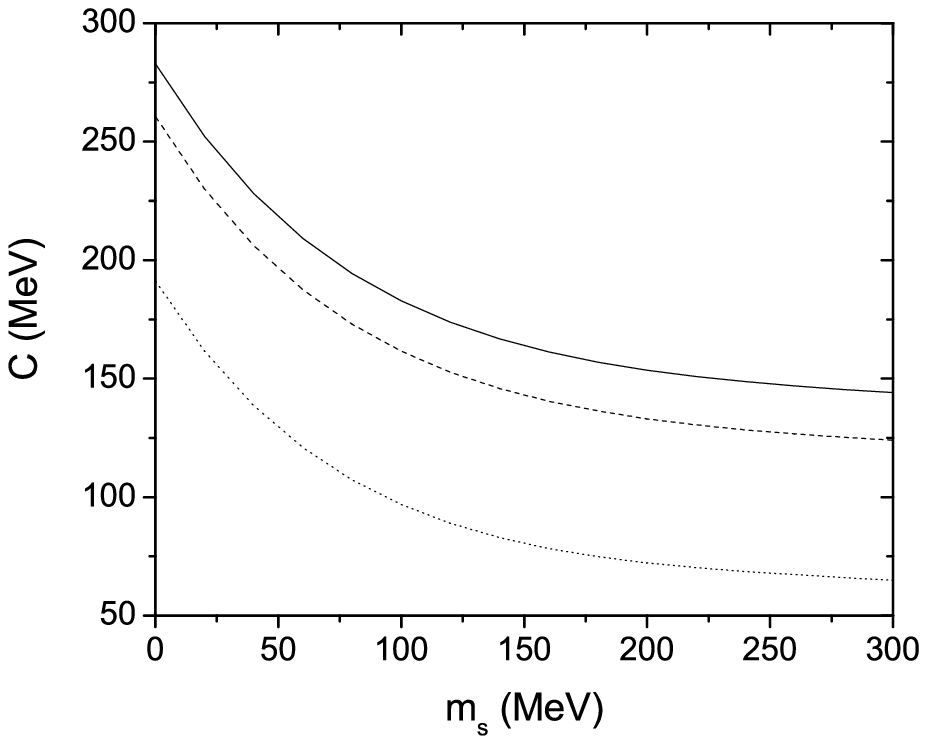}
\caption{Surface and curvature energies of CFL strangelets,
from left to right, respectively, as a function of $m_s$.
The value of the fixed constants are $T=0$ (full curve),
$T=15$ MeV (dashed curves), and $T=30$ MeV (dotted curve),
$A=100$ MeV, $B^{1/4}=145$ MeV, and $\Delta=100$ MeV.}\label{massa}
\end{figure}
\end{center}

\begin{center}
\begin{figure}
\includegraphics[width=0.45\textwidth]{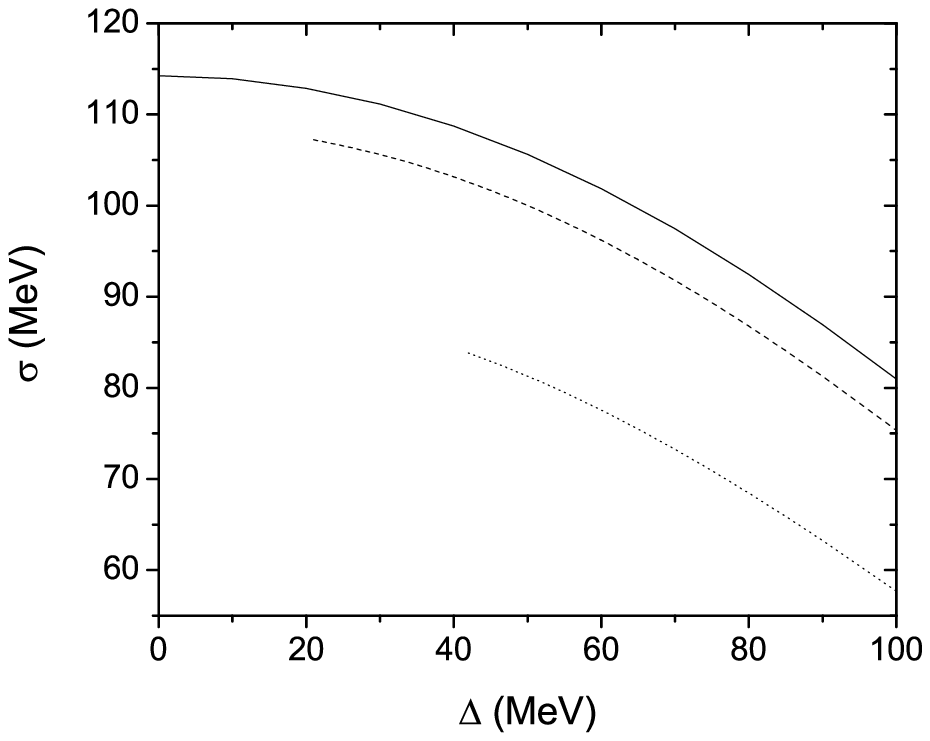}
\includegraphics[width=0.45\textwidth]{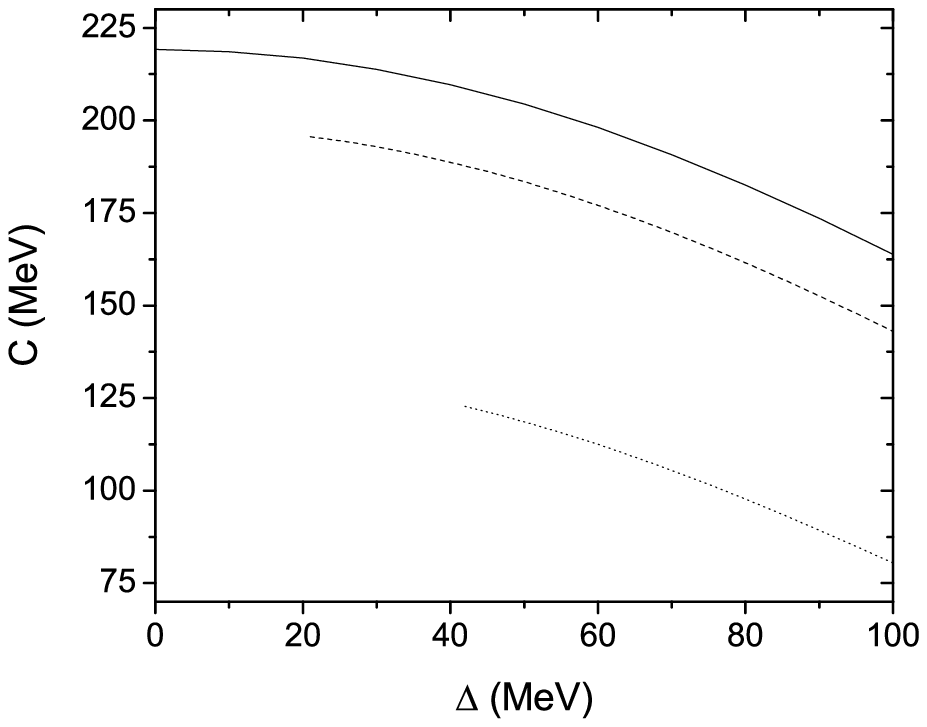}
\caption{Surface and curvature energies of CFL strangelets, from
left to right, respectively, as a function of $\Delta$. The value of
the fixed constants are $T=0$ (full curve), $T=15$ MeV (dashed
curves), and $T=30$ MeV (dotted curve), $A=100$, $m_s=150$ MeV,
and $B^{1/4}=145$ MeV. The critical temperature for
$\Delta\lesssim 20$ MeV is below fifteen MeV and for
$\Delta\lesssim 40$ MeV is below thirty MeV and that is why the
curves at finite temperature are plotted starting at different
values of the pairing gap energy.}\label{Delta}
\end{figure}
\end{center}

\end{widetext}

The behavior of the total energy per baryon number is to decrease
when increasing the pairing gap and the strangelet's baryon number,
and increase with increasing $B$, $m_s$ and $T$. These can be
understood with a comparison with the behavior of the stability
windows of SQM show in Fig. \ref{janela}.

The calculations show that the coefficient $R_0$, defining the
strangelet radius as in $R=R_0A^{1/3}$, decreases with increasing
$A$ and $B$ but increases with $m_s$ and the $\Delta$ parameter
(holding other parameters fixed on each comparison).
It is also higher whenever there is an increase in the temperature
for the thermic energy of quarks and gluons also increases.
These behaviors are easy to understand: increasing the strangelet's
baryon content, its parameters get closer to the bulk ones,
resulting in a decrease in $R/A^{1/3}$. Also, when increasing the bag
constant, the vacuum pressure on the strangelet's content is higher,
explaining the radius dependence on this parameter. With increasing
strange quark mass, the strange quark content decreases and so, with
fixed $A$, the radius increases to maintain the constraint
$A=\sum_iN_i$, the same reasoning applying to the raise in the pairing
gap.

In the following figures, the dependence of the surface and curvature
energies, defined as the coefficient that appears multiplying $A^{2/3}$
and $A^{1/3}$ in the expression for the total energy, respectively,
on these parameters is also shown.

The surface and curvature contributions decrease at higher
temperatures, a feature also seen in ordinary nuclear matter for
the surface energy (see, for example \cite{Ravenhall, Bondorf} and
references therein). The dependence of the surface energy with the
strange quark mass shows a maximum at $m_s\approx 150$ MeV and
goes to zero for massless quarks, and additionally shows a
decrease for high values of the strange quark mass due to a
depletion of this very massive component.

Simple numerical fits for the surface and curvature energy of
strangelets at finite temperature were also obtained (to
second order in the temperature $T$) for $\Delta=\Delta_0=100$ MeV
(that is, for a gap parameter independent of temperature),
$B^{1/4}=145$ MeV, and $m_s=150$ MeV and are presented here

\begin{widetext}
\begin{equation}
\sigma_{CFL}(T,A)=(81.09+0.013\,T-0.026\,T^2)\times(0.96+0.17\,e^{-\frac{A}{22.5}}+0.053\,e^{-\frac{A}{384.2}})
\end{equation}

\begin{equation}
C_{CFL}(T,A)=(163.85+0.003\,T-0.093\,T^2)\times(0.98+0.082\,e^{-\frac{A}{23.2}}+0.026\,e^{-\frac{A}{393.9}})
\end{equation}

\end{widetext}

As expected, the behavior of the parameters characterizing 
strangelets for CFL matter is qualitatively the same when compared
to SQM without pairing, i. e., the system at finite temperature is
less stable than at absolute zero but the surface and curvature
contributions decrease, a feature well known for nucleon systems.
One interesting point is that for $\Delta=100$ MeV, the chemical
potential for the $s$ quark is very close to the common chemical
potential for stable strangelets without pairing at the same
temperature and with the same value of $B$ and $m_s$, but the
chemical potential for the light quarks is much lower. As a
consequence the surface energy (determined only by the massive
strange quark) is almost equal for the two scenarios, but the
curvature energy is much lower for CFL strangelets. It means that
for values of the pairing gap lower than 100 MeV, the surface
energy is higher in the CFL state than without pairing. Meanwhile
the curvature energy is always lower for CFL strangelets,
regardless the value of $\Delta$.

\begin{center}
\begin{figure}
\includegraphics[width=0.45\textwidth]{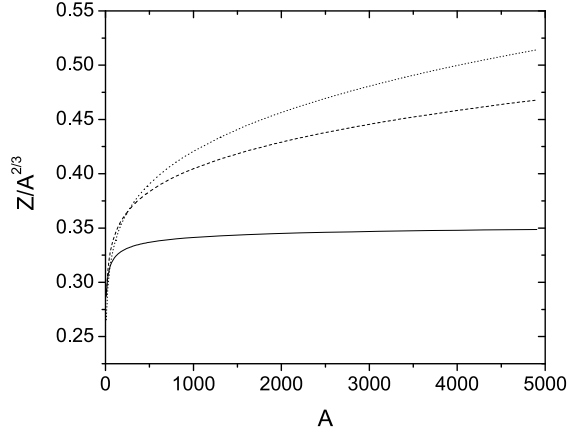}
\caption{Electric charge of strangelets as a function of the baryon number
for $T=0$ (full curve), $T=15$ (dashed curve), and $T=30$ (dotted curve), 
$B^{1/4}=145$ MeV, $m_s=150$ MeV, and $\Delta=100$.}\label{charge}
\end{figure}
\end{center}

The behavior of the electric charge for a strangelet with fixed
$T$, $B$, $m_s$, and $\Delta=\Delta_0$ as a function of the baryon
number is shown in Fig. \ref{charge}. The electric charge grows
with temperature for large baryon number strangelets driven by the
dependence on the number density of quarks with $T$. For the
massless quarks, a non-zero temperature slightly favor their
increase in number through the term $\mu T^2 V$. But for the
massive $s$ quark, the effect is the opposite since
$N_s=\int_0^{\infty} dNs/dk (1+\exp((\sqrt{k^2+m_s^2}-\mu_s)/T))
dk$. So $Z/A^{2/3}$ deviates from being close to a constant
behavior, expected from the suppression of massive strange quarks
near the surface at $T=0$ \cite{Madsen01}, due to the higher
importance of volumetric imbalance on the number density of
different quark flavors at higher temperatures.

\section{Conclusions}

CFL strange quark matter at $T=0$ is more stable than SQM without
pairing \cite{Madsen01,Lugones} when one considers the strange
quark mass, strong coupling constant and bag constant fixed. This
result has been extended and quantified for $T > 0$ in the
present work. Even when the temperature is as high as to be close
to the critical temperature for pairing there is still room for
(meta-) stability, depending on parameters choice. This suggests
that the process of transition from a neutron star to a strange
star could proceed right after its formation and the system might
even skip the neutron star stage, if conditions for conversion of
ordinary nuclear matter to the CFL state are met in the interior
of these compact objects. As a general result, it is possible to
observe that a finite temperature has always the consequence of
destabilizing the system even when considering the dependence of
$\Delta$ with $T$ which causes SQM to be slightly more
disfavored than its ``constant $\Delta$'' version.

We also notice a very distinctive feature between strangelets with
and without pairing concerning the existence of the critical
baryon number, $A_{crit}$. This quantity represents the minimum
baryon number to which strangelets are stable against neutron
decay. The effects of surface and curvature tend to destabilize
strange matter at low baryon number. As a result, the energy for
creating small lumps of SQM increases as the baryon number
decreases till it reaches a value above the neutron decay
threshold, i. e., till it is above $\sim 930$ MeV. As shown in
\cite{Chineses} and \cite{Madsen98}, the critical baryon number
exists even for null temperature. It is known (see \cite{Farhi84}
and \cite{Lugones}) that the lower the value of the bag constant
(of course, respecting the limit $B^{1/4}\geq 145$ MeV) and the
higher the value of the pairing gap, the more stable strange quark
matter in bulk is. In the case of CFL strangelets with high values
of $\Delta$ and relatively low values of the bag constant,
performing the analysis within the MIT bag model, the existence of
$A_{crit}$ is not clear. It must be noted, however, that the
liquid drop model does not provide a good description for low
baryon number, being less reliable than shell models filling
explicitly the quark states \cite{Jaf}.

Another important feature is that although for high baryon number
CFL strangelets seem to be absolutely stable even for temperatures
of order 30 MeV when $\Delta=100$ MeV, values of the pairing
constant much above a few hundreds of MeV are not expected to
describe these systems. In being so, the critical temperatures for
pairing of quarks inside strangelets are not expected to be higher
than 70-100 MeV. For temperatures above this value, the quarks
inside the strangelet would no longer be paired and the gain in
energy for this state compared to non-superconducting strangelets
would vanish. Since strangelets without pairing are not stable for
temperatures as high as the maximum critical temperatures
expected, the strange quark matter stability would vanish above
$T_{crit}$ if the pairing gap is too high.

We have also found that strangelets are more favorable in the CFL
state, as expected. In particular, the curvature energy for CFL
strangelets is lower than for ``normal'' strange quark matter,
which may influence the fragmentation process of bulk CFL SQM, an
important issue when considering the possible presence of these
particles among cosmic rays and also when considering strangelet
production in heavy ion collisions, although the very high
temperatures disfavor the production of stable strangelets in
these environments \cite{Madsen01}.

\begin{acknowledgements}
We acknowledge the very important advice of M. Alford on the
issue of screening for CFL strangelets. This work is supported by
Funda\c c\~ao de Amparo \`a Pesquisa do Estado de S\~ao Paulo. JEH
acknowledges the partial financial support of the CNPq Agency
(Brazil).
\end{acknowledgements}

\bibliography{CFLT}

\end{document}